\documentclass[proceedings, preprint]{rmaa}

\usepackage{paralist}

\usepackage{psfrag,color}

\SetYear{2024}
\SetConfTitle{}

\title{Espartaco 2, a new stellar spectrograph at Uniandes} 

\author{
  B. Oostra,\altaffilmark{1} 
  M.G. Batista,\altaffilmark{1}}

\altaffiltext{1}{Universidad de los Andes, Departamento de Física, Carrera 1 \#18a - 12, 111711, Bogotá, Colombia, (boostra@uniandes.edu.co), (mg.batistar@uniandes.edu.co).}

\shortauthor{Oostra \& Batista}
\shorttitle{Espartaco 2, a new stellar spectrograph}

\listofauthors{B. Oostra, M.G. Batista}
\indexauthor{Oostra, B.}
\indexauthor{Batista, M.G.}

\abstract{We present the construction and early results of ESPARTACO 2, the new stellar spectrograph built for research and education at the Astronomical Observatory of the Universidad de los Andes in Bogotá, Colombia. This instrument offers several resolutions from 20,000 in first order using a 50 $\mu$m fiber, to 100,000 in second order in the near infrared. Precise radial-velocity measurements are made possible by simultaneous wavelength calibration. Combined with the 40-cm Meade telescope located at our facilities, a limiting magnitude of 6 is reached. This instrument is a considerable improvement over its predecessor in throughput, reliability and ease.}

\resumen{Presentamos la construcción y los primeros resultados de ESPARTACO 2, el nuevo espectrógrafo estelar que fue construido con fines educativos y de investigación en el Observatorio Astronómico de la Universidad de los Andes en Bogotá, Colombia. Este instrumento tiene la capacidad de ofrecer varias resoluciones, desde 20,000 cuando se usa la fibra de 50 $\mu$m en primer orden, hasta 100,000 en el segundo orden en el infrarojo cercano. Es posible realizar medidas precisas de velocidad radial estelar mediante la calibración simultánea en longitud de onda. Al combinarse con el telescopio Meade de 40 cm localizado en nuestras instalaciones, se logra estudiar objetos hasta magnitud 6. Este instrumento representa una mejora respecto a su antecesor en cuanto a su eficiencia lumínica y facilidad de uso.}

\addkeyword{Spectroscopy}
\addkeyword{Instrumentation}
\addkeyword{Radial Velocity}
\addkeyword{Education}

\begin{document}
\maketitle

\section{Introduction}
\label{sec:intro}

Our campus observatory rose from the need of making observations that could complement the theoretical astronomy courses. Our location near Bogotá's city center has some drawbacks such as light pollution, severe seeing and very variable weather, but the advantage is that we can seize the opportunity when conditions are good.

In this urban environment, high-resolution spectroscopy of the sun and bright stars is more promising than any other astronomical technique, and has proven a fruitful endeavor \citep[]{Oostra:2012, Oostra:2017, Oostra:2024} through 15 years of work with \textit{ESPARTACO, Espectrógrafo de Alta Resolución para Trabajos Astronómicos en Colombia} \citep{Oostra:2011}. This instrument is a high-resolution, low-cost spectrograph designed for undergraduate projects. Pursuing this path, half a dozen spectrographs have been built and employed at our Observatory over the past 3 decades. 

Despite all the goals achieved, ESPARTACO has several problems. It was built in 2007 on a non-optimal design, from parts that were at hand at the moment; the mismatch between these components produces a very low throughput. Moreover, wavelength calibration spectra must be taken before or after the science spectra, mechanical stability is deficient, and proper focusing is a real challenge. 

Aside from its difficulties, ESPARTACO has provided valuable experience in astronomical studies as well as a laboratory for atomic physics. This justified some investment in more optimized optical components, prompting the construction of a new instrument that began in 2018 and finished in 2023. Building the instruments locally is a way of cost reduction, but also provides experience in astronomical instrumentation. In this paper we share in Section 2 a technical description of the instrument. In section 3 we show some results from quality test. Finally, section 4 provides some final remarks.

\section{Description of the instrument}
\label{sec:desing}

\subsection{Optical layout}

\begin{figure}[!t]
  \includegraphics[width=\columnwidth]{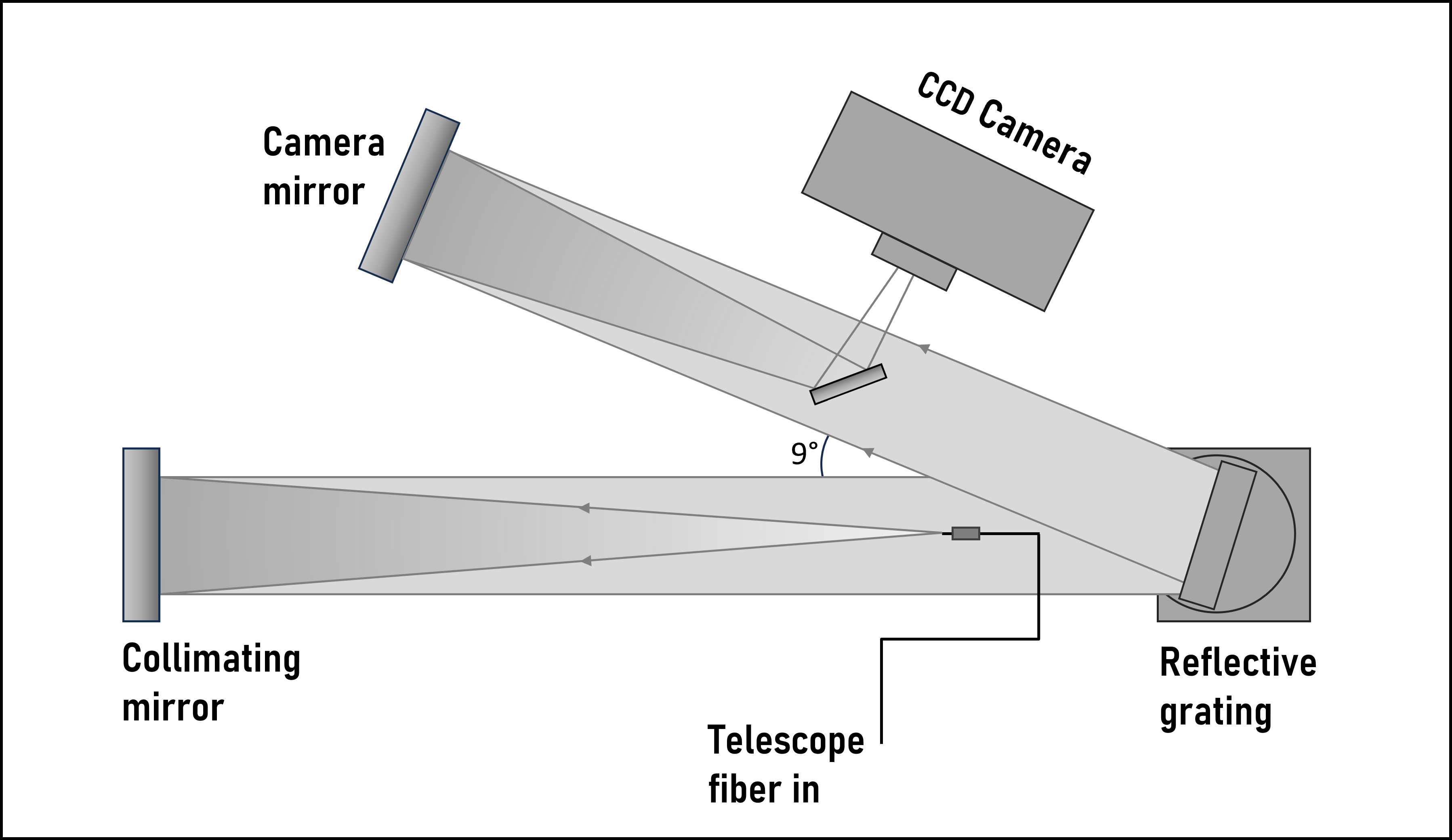}
  \caption{Optical layout of ESPARTACO 2. The angle between the collimator and the camera is 9°. The only moving parts are the grating, which can rotate 360°, and the fiber support. }
  \label{fig:layout}
\end{figure}

ESPARTACO 2 is a Czerny-Turner spectrograph with on-axis parabolic mirrors of 15 cm diameter. The grating is 14 cm wide and 12 cm high and has a density of 1200 grooves/mm without blaze. The collimator length is 120 cm (F/8) and the camera mirror has 76 cm (F/5). The instrument can be used in first and second diffraction order, see Figure~\ref{fig:layout}.

\subsection{Fibers}

The instrument is fed via 4 optical fibers of circular cross-section: three of 50 $\mu$m diameter and one of 10 $\mu$m. At the collimator-end of these fibers, their tips were polished without ferrules, in order to minimize the spacing between them; their plastic sleeves keep the cores at 0.9 mm from each other, which places the images in the camera at a vertical spacing of 63 pixels. The exit beam of these fibers is wider than the F/8 collimator, resulting in some loss of luminosity. The multiplicity of fibers makes it possible to register simultaneously the calibration spectrum on the same image as the stellar spectrum; this is necessary for precise measurements of radial velocities, and was not possible with our previous spectrograph. For stellar observations, one 50 $\mu$m fiber carries the light from the star, another brings the light from a Th-Ar hollow-cathode lamp, and a third one may be coupled to a neon lamp for eventual fast position or focus checks. Obviously, there is a small horizontal offset between any pair of fibers, which must be measured empirically. The star fiber is coupled to the telescope via an interface from Shelyak Instruments\footnote{In order to reduce the size of the stellar image in the telescope's focal plane, and thus maximize the amount of light though the optical fiber, we connect to the telescope a Meade Electronic Micro-Focuser, Meade series 4000 f/6.3 focal reducer and the Shelyak FIGU (Fiber Injection and Guiding Unit) to inject the starlight into an optical fiber and a guiding camera.}.

\subsection{Camera}

The camera is a Kodak KAF1603ME CCD array of 1530 x 1020 pixels measuring 9 $\mu$m square. It is commonly operated at –15$^\circ$C. Downloading the images via a USB cable takes a few seconds. Some technical specifications:

\begin{itemize}
    \item Full well = 100000 electrons, converted to 65535 ADU, give a gain of 1.52 e/ADU
    \item Read-out noise: 10.5 ADU (RMS) = 16 electrons.
    \item Dark current (reported):  1 electron per pixel per second at 0$^\circ$C
    \item Dark current (measured): 1 electron per pixel per minute at –$15^\circ$C
    \item Quantum Efficiency: 75\% at H-$\alpha$, 50\% at H-$\beta$.
    \item Sensivity range: 400 - 800 nm.
\end{itemize}

\subsection{Structure}

The optical components are mounted in a rigid 165 x 45 x 35 cm steel frame, made of 50-mm-wide steel angle, covered with black acrylic panes. The only moving parts are the grating’s support, which can rotate 360$^\circ$, and the fiber support which can travel 2 cm for focusing. Both these devices are driven by small stepper motors which are controlled manually. The angular position of the grating is monitored by a webcam. The stepper motors and the controlling webcam are important improvements over the previous version of ESPARTACO 1 \citep{Oostra:2011}.

\subsection{Thermal instability}

A major problem is the lack of a thermally stabilized room for the spectrograph. The instrument is located in the small dome building where the temperature may change from below 10$^\circ$C at 3:00 a.m. to more than 20$^\circ$C at 3:00 p.m. This poses the need of adjusting the focus. But extreme refocusing will degrade the spectral line profile, so we try to mitigate the thermal changes using insulation (20 mm Ethafoam) and a low-power internal heating system.

\section{Results}

\subsection{Resolution}

To assess the resolution of the spectrograph we performed a series of line width measurements on carefully focused spectra of our Th-Ar lamp. We used only Thorium lines because these are perceptibly narrower than Argon lines, possibly due to the greater atomic mass which reduces Doppler broadening.

The measurements were made using the same 50 $\mu$m fibre that is commonly employed for stars.

On each profile we adjusted a Gaussian fit to the 4 or 5 central data points. The reported line widths are the FWHM of these fits. Each selected Th line was measured in first and in second order.

\begin{figure}[!t]
  \includegraphics[width=\columnwidth]{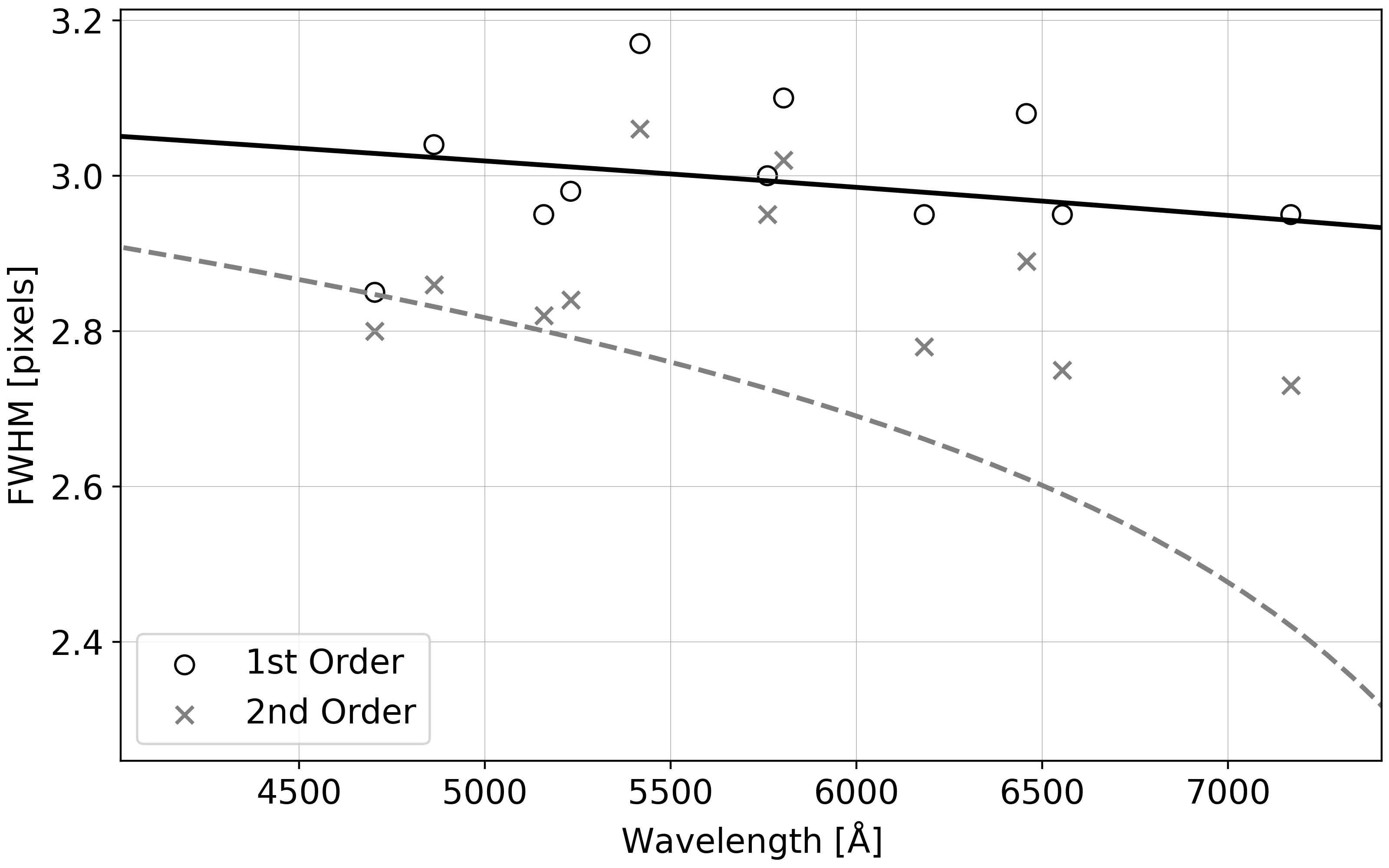}
  \caption{Measured widths of Th emission lines. Circles and crosses represent first order and second order measurements, respectively. Theoretical line widths are shown with solid line for the first order and dashed for second order.}
  \label{fig:widths}
\end{figure}

\begin{figure}[!t]
  \includegraphics[width=\columnwidth]{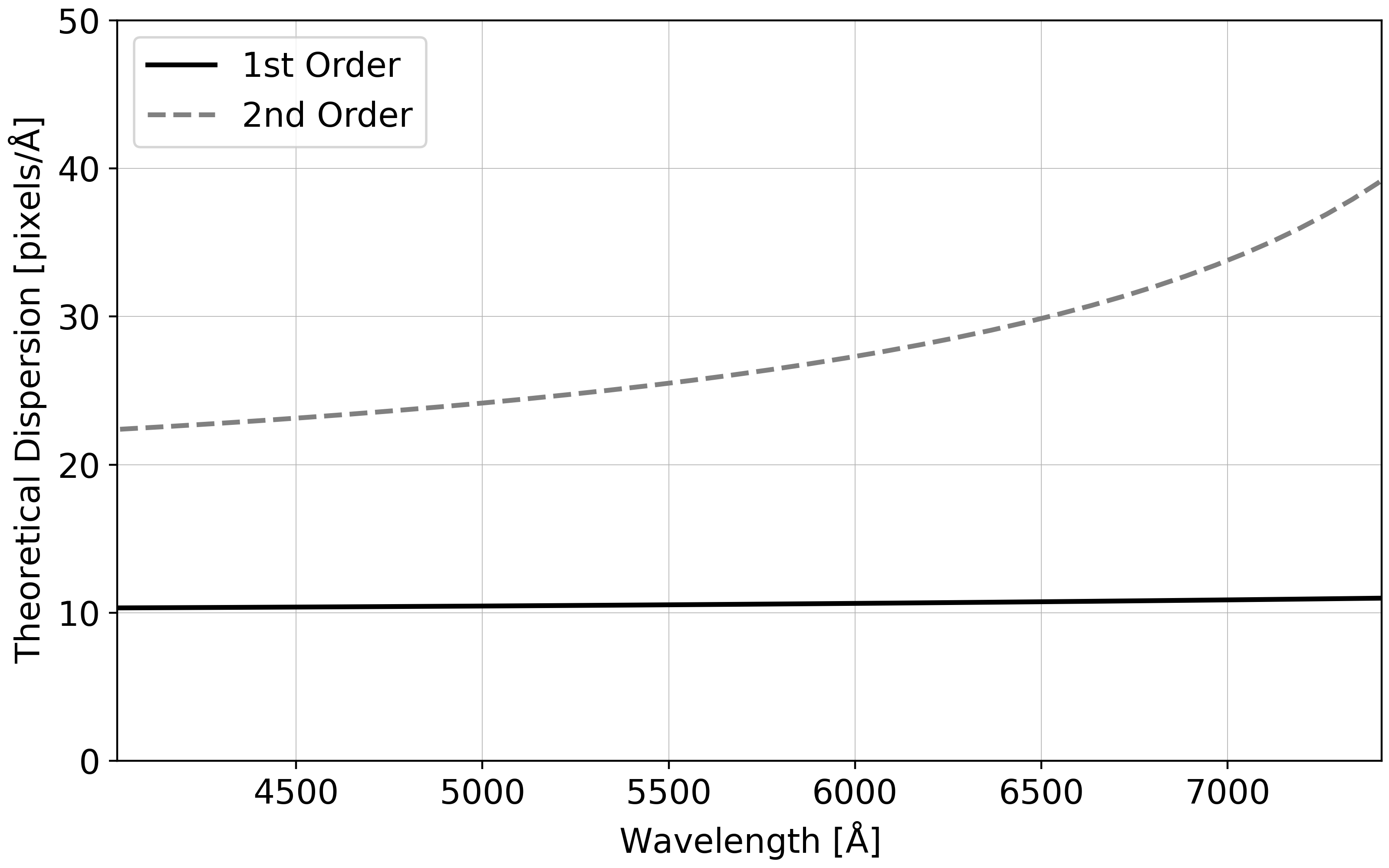}
  \caption{Theoretical dispersion of the spectrograph. Solid line for first order and dashed for second order. Observed values are found to agree with the theory.}
  \label{fig:dispersion}
\end{figure}

\begin{figure}[!t]
  \includegraphics[width=\columnwidth]{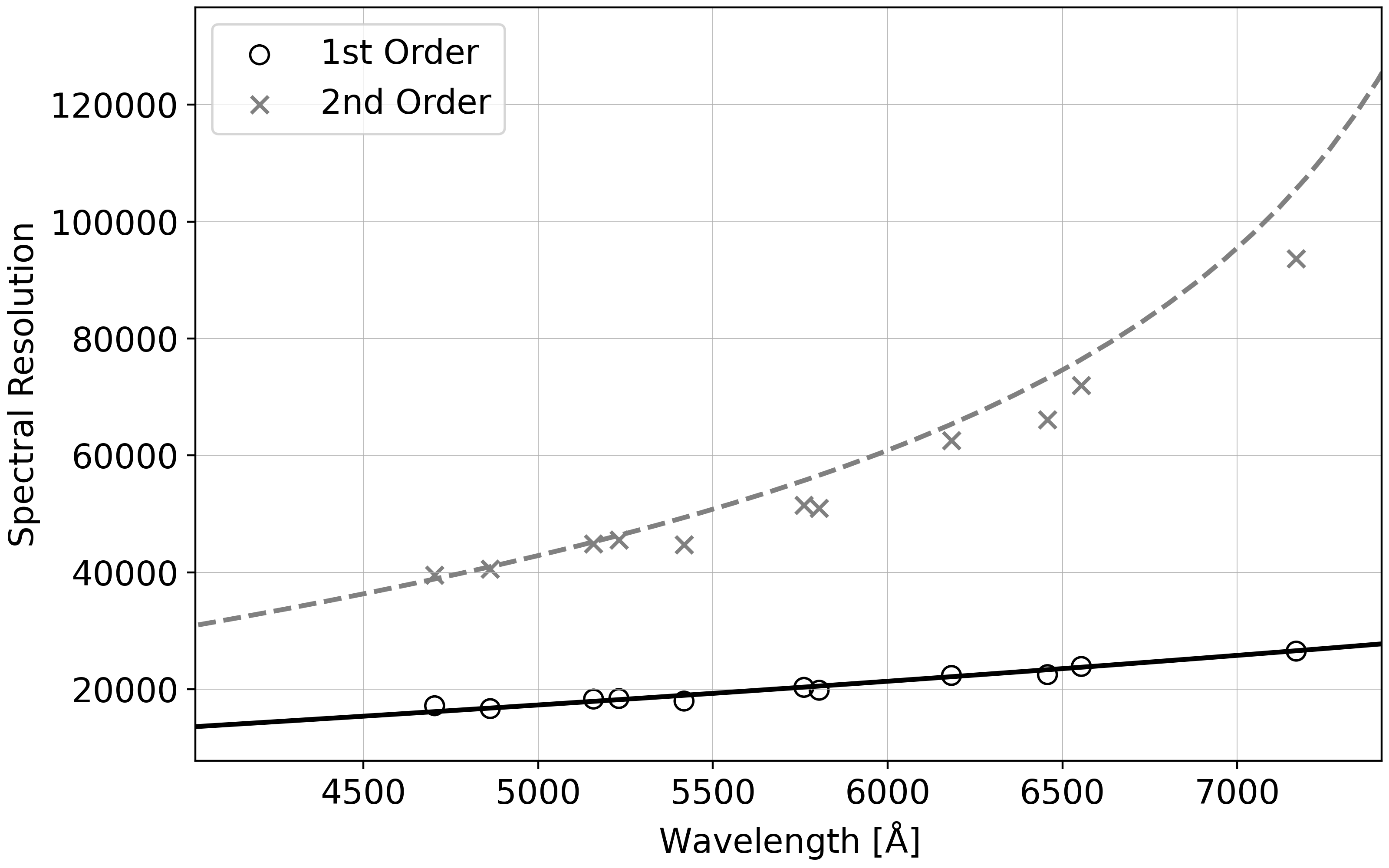}
  \caption{Spectral resolution computed from dispersion and line width. Conventions as before.}
  \label{fig:resolution}
\end{figure}

Figures~\ref{fig:widths}, ~\ref{fig:dispersion}, ~\ref{fig:resolution} show the theoretical and measured line widths, the theoretical dispersion, and the theoretical and measured resolution. For the calculations we include only geometrical optics, no diffraction; and instead of a circular input, we assumed a rectangular entrance slit of 45 micrometers. As can be seen, stars must be observed at resolution 20.000 in first order, which is considerably less than the 31.000 given by our previous spectrograph.  

Through the narrow 10 $\mu$m fibre we took some images of the particularly strong Th 5761 line. In first order a width of 2.03 pixels can be readily measured, suggesting that this fibre enhances the resolution by about 50\% for bright emission lines in first order. In second order the focusing is very ambiguous and the results are not clear: in several images it is possible to fit a gaussian to the 3 central points and obtain a width of 1.8 or 1.9 pixels, but a great amount of light is scattered outside this peak. This shows that diffraction effects are not negligible. A line width below the Nyquist limit of 2 pixels is not very useful anyway.



\begin{figure}[!t]
  \includegraphics[width=\columnwidth]{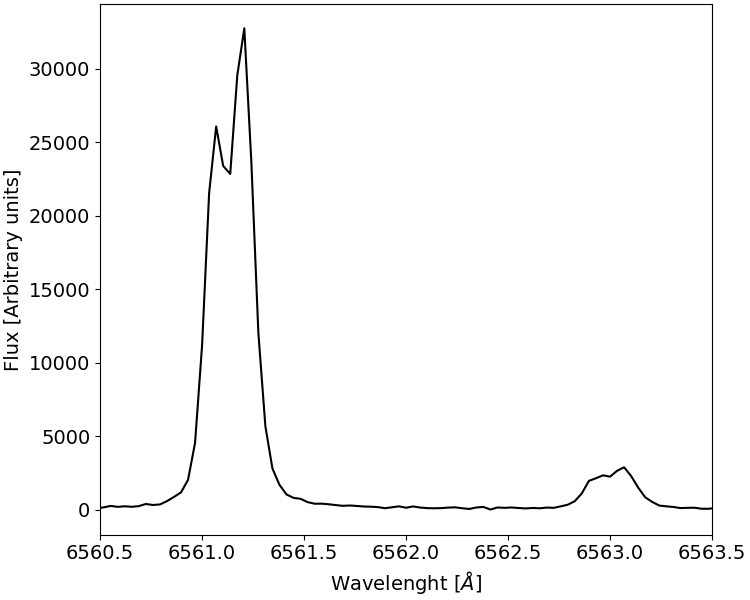}
  \caption{Spectrum of deuterium-$\alpha$ and its fine structure, obtained with 20 seconds of exposure using the 10 $\mu$m fiber in second order (left). A trace of normal hydrogen, which is present in the deuterium tube, produces the small emission line at right. The isotopic shift constitutes a regular student assignment. For this experiment, deuterium is better than normal hydrogen because it has less thermal broadening.}
  \label{fig:deu}
\end{figure}

The light transmitted by the 10-$\mu$m fiber is very weak, and exposure times would be prohibitively long. And on the other hand, continuous spectra through this fiber are modulated by intermodal interference \citep[]{Hlubina:1999, Baudrand:2001, lemke:2011}. Nevertheless, it is useful for laboratory spectra, to study, for example, the Zeeman effect and the fine structure of deuterium (Figure \ref{fig:deu}).


\subsection{Throughput and limiting magnitude around H-alpha}

The main improvement of ESPARTACO 2 over our previous spectrograph is the gain in luminosity; this was also the main reason for building a new instrument. The new one has 14 times greater optical throughput than its forerunner, as measured by aperture photometry on neon spectra. This represents a progress of almost 3 stellar magnitudes, and is mainly due to the size of the new grating, 120 x 140 mm as compared to the old grating’s 50 x 50 mm.

\begin{figure*}[!t]
  \includegraphics[width=2.0\columnwidth]{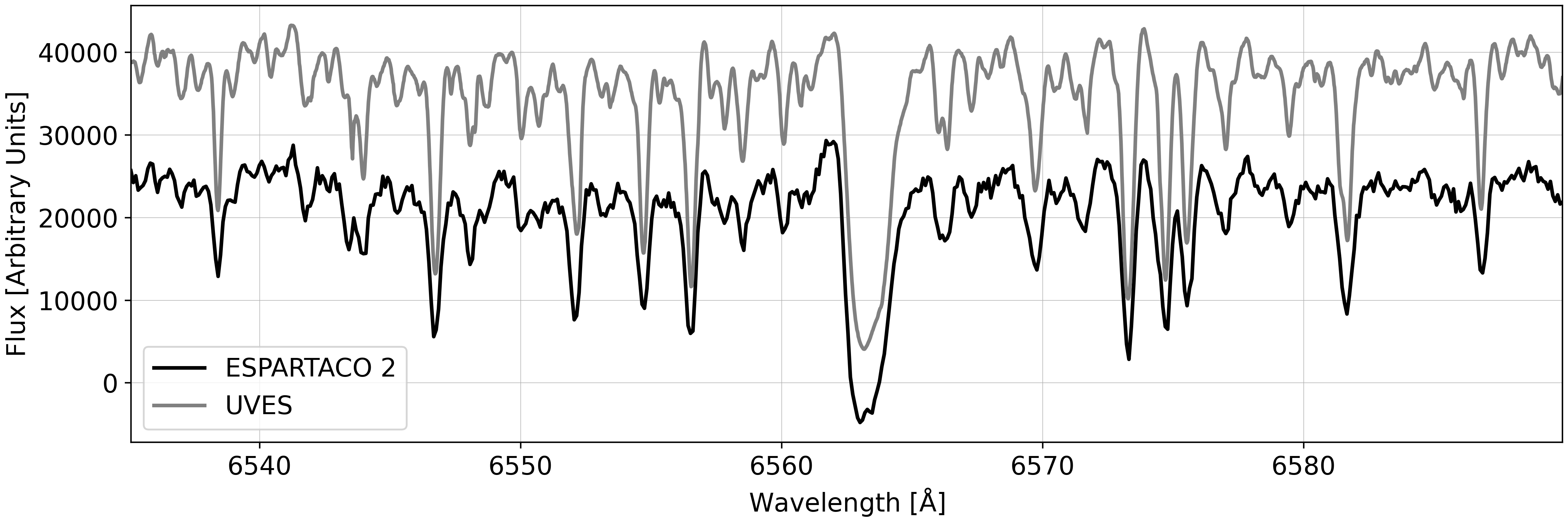}
  \caption{Spectrum of HD49331 obtained with ESPARTACO 2 (black line); for comparison, the same spectrum from UVES (grey line), at resolution 80000; \citep{UVES:2003}.}
  \label{fig:HD49331}
\end{figure*}

The limiting magnitude depends somewhat on the stellar spectral type and on the observed wavelength range. If we define the limit by a S/N ratio of 10 and an exposure time of 20 minutes on our 40-cm telescope, the limiting V magnitude is about 5 for early-type stars and 6 for late-type stars observed at H-alpha in first order. As an example, Figure~\ref{fig:HD49331} shows a spectrum of HD49331, a red supergiant of magnitude V = 5.1, obtained with a 20-minute exposure and giving a signal-to-noise ratio of 38.

Figure~\ref{fig:HD49331} compares our spectrum of HD49331 with a spectrum of the same star recorded at resolution 80.000 by the UVES spectrograph and published under the Paranal Observatory Project \citep{UVES:2003}. The comparison shows that the main features are clearly legible in our data, and also many smaller features that we are missing on account of our lower resolution and signal strength.

\begin{figure*}[!t]
  \includegraphics[width=2.0\columnwidth]{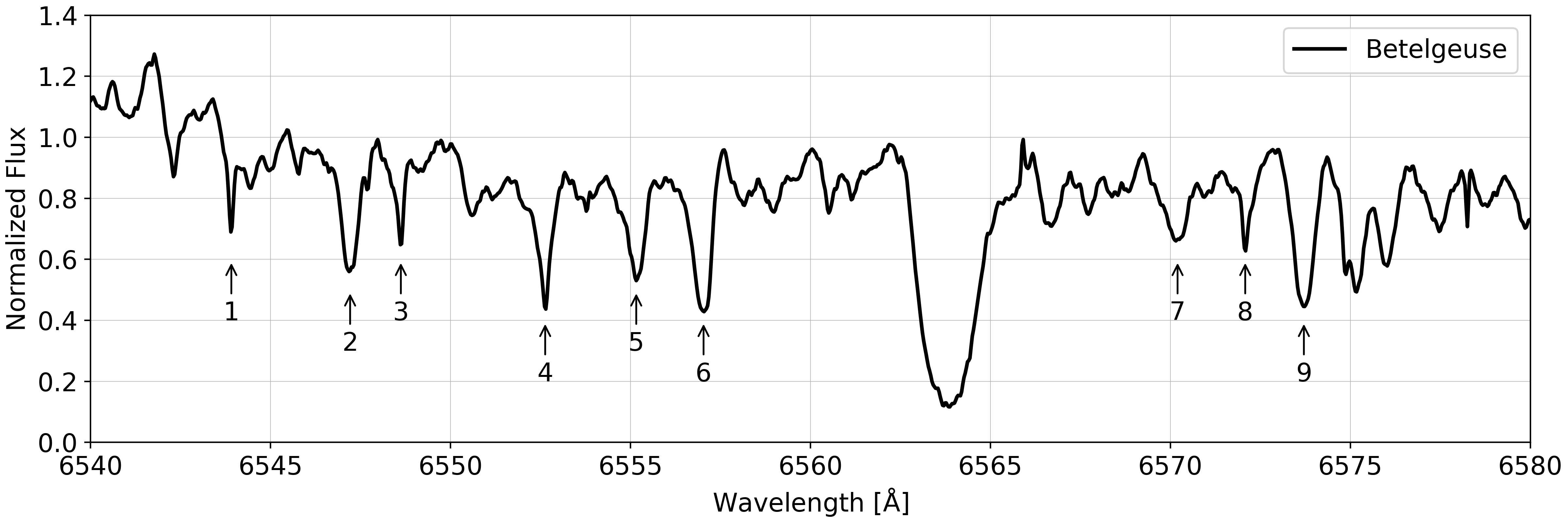}
  \caption{Spectrum of Betelgeuse obtained with ESPARTACO 2 in second order in a 10-minute exposure. The numbers 1 to 9 mark selected absorption lines discussed in the text (Table~\ref{Tab:lines}). In particular, lines 1, 3, 4 and 8 are telluric H$_2$O lines, and are evidently narrower than the stellar lines; this distinction is not visible in the first-order spectrum shown in Figure~\ref{fig:HD49331}.}
  \label{fig:Betelgeuse}
\end{figure*}

Besides the Sun, we can study some bright stars in second order. As an example, Figure~\ref{fig:Betelgeuse} gives the second-order spectrum of Betelgeuse obtained in a 10-minute exposure. In this configuration, the higher resolution allows some line width measurements; the results are shown in Table~\ref{Tab:lines}. Most conspicuous is that stellar lines are broader than telluric lines, due to the star’s turbulence and rotation. This distinction is not visible in Figure~\ref{fig:HD49331}, partly due to the lower resolution at first order, and also because HD49331 is less turbulent.

Second-order spectra have the disadvantage that they span a range of only 50 \AA, while first-order spectra cover 140 \AA. Moreover, exposure times must be long, and the limiting magnitude is about 1. This option is particularly useful for solar and laboratory studies.

\begin{table}[]
\caption{prominent absorption lines in the Betelgeuse spectrum}


\begin{tabular}{|l|l|l|l|l|l|}
\hline
\multicolumn{1}{|c|}{\#} & \multicolumn{1}{c|}{$\lambda$ obs}            & \multicolumn{1}{c|}{Species} & \multicolumn{1}{c|}{Width}                    & \multicolumn{1}{c|}{$\lambda$ rest}           & \multicolumn{1}{c|}{RV}         \\
\multicolumn{1}{|c|}{}      & \multicolumn{1}{c|}{[\AA]} & \multicolumn{1}{c|}{}        & \multicolumn{1}{c|}{[\AA]} & \multicolumn{1}{c|}{[\AA]} & \multicolumn{1}{c|}{[km/s]} \\ \hline
1                           & 6543.91                                       & H$_2$O                       & 0.15                                          & 6543.907                                      & 0.14                            \\ \hline
2                           & 6547.21                                       & Fe+Ti                      & 0.48                                          & 6546.253                                     & 43.86                           \\ \hline
3                           & 6548.62                                       & H$_2$O                       & 0.18                                          & 6548.622                                      & -0.09                           \\ \hline
4                           & 6552.63                                       & H$_2$O                       & 0.22                                          & 6552.629                                      & 0.05                            \\ \hline
5                           & 6555.16                                       & Ti                           & 0.52                                          & 6554.223                                      & 42.89                           \\ \hline
6                           & 6557.03                                       & Ti                           & 0.57                                          & 6556.062                                     & 44.30                           \\ \hline
7                           & 6570.20                                       & Fe+Fe                        & 0.65                                          & 6569.215                                     & 45.00                           \\ \hline
8                           & 6572.08                                       & H$_2$O                       & 0.18                                          & 6572.086                                      & -0.27                           \\ \hline
9                           & 6573.71                                       & Ca                           & 0.65                                          & 6572.779                                      & 42.49                           \\ \hline
\end{tabular}
\label{Tab:lines}
\end{table}

\subsection{Budget}

All optical components were purchased off-the-shelf. The 15-cm mirrors and supports are from Edmund Optics, as well as the diagonal mirror. Both motor stages are from Optimal Engineering Systems, Inc. (OES). The grating is from Horiba Jobin-Yvon. The camera is from SBIG (taken over by Diffraction Limited). The prices of these components add up to about 17000 USD.

The structural frame, the mounting for the diagonal mirror, as well as some other parts, were built at our Physics Department Mechanics Laboratory. Optical fibers and other hardware were purchased at low cost on the local market. We believe that the overall cost/benefit ratio is reasonable, considering that our environmental conditions do not justify high-cost equipment.

\section{Final remarks}

ESPARTACO 2 will be useful for research and teaching from high school to graduate level. This instrument has the capacity to offer several resolutions, depending the order and the diameter of the optical fiber used, has a greater luminous throughput and the advantage of moving parts with stepper motors, delivering considerable improvements over its predecessor. However, eventual refinement and reformations may be implemented in the future:

\begin{itemize}
    \item Add another 50 $\mu$m fiber with a micro-lens in order to couple the fiber optimally to the collimator aperture. This would upgrade the limiting magnitude, albeit with lower spectral resolution.
    \item Install a chiller for water-cooling the camera.
    \item Couple the focus motor and heating system to a thermostat.
    \item Implement digital encoders for the position of the grating and the focus motor.
    \item Replace the mirrors by off-axis paraboloids, which would eliminate the diagonal mirror and place the fibers outside the collimator beam, enhancing the image quality and luminosity.    
\end{itemize}

\end{document}